\documentclass[
    ,final            
  ]
  {aipproc}

\layoutstyle{6x9}

\newcommand{\mean}[1]{\mbox{$\langle#1\rangle$}}
\newcommand{\scint}[2]{\mbox{$#1\!\times\!10^{#2}$}}

\newcommand{\gtrsim} {\stackrel{\textstyle{>}}{\sim}}
\newcommand{\lesssim}{\stackrel{\textstyle{<}}{\sim}}

\begin{document}

\title{Fluctuations and Pinch-Offs Observed in Viscous Fingering}

\author{Mitchell G.\ Moore}
{address={Center for Nonlinear Dynamics and Department of Physics,\\
The University of Texas at Austin, Austin, Texas, 78712}}

\author{Anne Juel}
{address={Center for Nonlinear Dynamics and Department of Physics,\\
The University of Texas at Austin, Austin, Texas, 78712}
,altaddress={Present address: Dept. of Mathematics,
\\University of Manchester, Manchester M13 9PL, United Kingdom}}

\author{John M.\ Burgess}
{address={Center for Nonlinear Dynamics and Department of Physics,\\
The University of Texas at Austin, Austin, Texas, 78712}}

\author{W.\ D.\ McCormick}
{address={Center for Nonlinear Dynamics and Department of Physics,\\
The University of Texas at Austin, Austin, Texas, 78712}}

\author{Harry L.\ Swinney}
{address={Center for Nonlinear Dynamics and Department of Physics,\\
The University of Texas at Austin, Austin, Texas, 78712}}

\begin{abstract}

Our experiments on viscous (Saffman-Taylor) fingering in Hele-Shaw
channels reveal several phenomena that were not observed in
previous experiments.  At low flow rates, growing fingers undergo
width fluctuations that intermittently narrow the finger as they
evolve.  The magnitude of these fluctuations is proportional to
$\mathrm{Ca}^{-0.64}$, where Ca is the capillary number, which is
proportional to the finger velocity.  This relation holds for all
aspect ratios studied up to the onset of tip instabilities. At
higher flow rates, finger pinch-off and reconnection events are
observed. These events appear to be caused by an interaction
between the actively growing finger and suppressed fingers at the
back of the channel. Both the fluctuation and pinch-off phenomena
are robust but not explained by current theory.

\end{abstract}

\maketitle



Viscous fingering occurs when a less viscous fluid displaces a more viscous
fluid in a Hele-Shaw channel (a quasi-2D geometry in which the width $w$ is much
greater than the channel thickness $b$); the interface between the fluids is
unstable and forms a growing pattern of ``fingers''.  A single finger forms at
low flow rates; more complex branched patterns evolve at high flow rates.  This
phenomenon is the simplest example of the class of interfacial pattern forming
systems which includes dendritic growth and flame propagation.  Viscous
fingering thus continues to receive attention for the insight it provides into
these important
problems~\cite{Bensimon/Kadanoff/etc:1986,Homsy:1987,DynCurvedFronts,Couder:1991}.

Saffman and Taylor first studied the problem in 1958~\cite{Saffman/Taylor:1958}
by injecting air into oil in a Hele-Shaw cell. They observed the formation of a
single, steadily moving finger whose width decreased monotonically to $1/2$ of
the channel width as the finger speed was increased. Subsequent
experimental~\cite{Tabeling/Zocchi/etc:1987},
numerical~\cite{DeGregoria/Schwartz:1986}, and
theoretical~\cite{McLean/Saffman:1981,McLean:1980} work found that the ratio
of finger width to channel width, $\lambda$ depended on a modified capillary
number, $1/B=12\,(w/b)^2\,\mathrm{Ca}$, which combines the aspect ratio, $w/b$,
and the capillary number, Ca~$=\mu V/\sigma$, where $\mu$ is the dynamic
viscosity of the liquid, $V$ is the velocity of the tip of the finger, and
$\sigma$ is the surface tension.  A transition to
complex patterns of tip-splitting occurs at large $1/B$ values%
~\cite{Tabeling/Zocchi/etc:1987,Park/Homsy:1985,Kopf-Sill/Homsy:1988b,Maxworthy:1987}.
Our experiments have revealed two phenomena that were not reported in
prior experiments%
~\cite{Saffman/Taylor:1958,Tabeling/Zocchi/etc:1987,Park/Homsy:1985,Kopf-Sill/Homsy:1988b}
or predicted theoretically: fluctuations in the width of the
evolving viscous fingers~\cite{Moore/Juel/etc:2002}, and finger
pinch-off events.


\section{Experimental Methods and Data Analysis}

We conducted experiments in a 254~cm long channel formed from 1.9~cm thick glass
plates. The spacing between the glass plates was set by stainless steel strips
with thicknesses $b=0.051$~cm, 0.064~cm, 0.102~cm, or 0.127~cm; the channel
width $w$ between the spacers was varied between 19.9~cm and 25.1~cm. Both glass
plates were supported and clamped at the sides, and the channel was illuminated
from below. For aspect ratios under 150, we used a smaller channel of length
102~cm and width 7.4~cm. Interferometric measurements revealed that the
root-mean-square variations in gap thickness were typically 0.6\% or less in the
large channel and 0.8\% in the small channel.  Mechanical measurements of the
bending of the glass due to the imposed pressure gradient revealed that such
deflections were typically 0.2\% or less; the maximum deflection (2.2\%) was
measured in the widest channel close to the oil reservoir at the highest flow
rates.  Experiments were conducted with air penetrating a Dow Corning silicone
oil whose surface tension and dynamic viscosity was either
$\sigma=19.6$~dyne/cm, $\mu=9.21$~cP or $\sigma=20.6$~dyne/cm, $\mu=50.8$~cP at
laboratory temperature ($22^\circ$C). The oils wet the glass completely.  We
withdrew oil at a uniform rate using a syringe pump attached to a reservoir at
one end of the channel; an air reservoir at atmospheric pressure was attached to
the other end. The entire experiment was placed on a floating optical table to
minimize vibrations and allow precise levelling of the channel.

We obtained images of up to $1200 \times 10,000$ pixels at a resolution of 0.25
mm/pixel using a camera and a rotating mirror.  The camera captured up to 11
overlapping frames which were then concatenated, background subtracted, and
corrected for perspective effects.  The interfaces were then digitally traced,
yielding finger width values accurate to 0.1\% in the larger channel and 0.3\%
in the smaller channel. For each flow rate, up to four time sequences of 20-30
digital interfaces were recorded. Finger widths determined in consecutive
sequences agreed within the measurement accuracy. Mean width values agreed
within 0.5\% for data sets repeated after channel disassembly, cleaning, and
reassembly. For each experimental run, we determined the time average
$\mean{\lambda}$ and the root-mean-square (rms) fluctuation from the mean
$\delta_\lambda$ (as described in~\cite{Moore/Juel/etc:2002}). Each data set was
analyzed for flow rates up to the point of tip splitting, beyond which the
finger width $\lambda$ was no longer well defined~\footnote{The first tip
instabilities observed with increasing $1/B$ are asymmetric ``shouldering''
modes, not actual tip
splitting~\cite{Bensimon/Kadanoff/etc:1986,Homsy:1987,DynCurvedFronts,Couder:1991,Tabeling/Zocchi/etc:1987};
because the first instabilities are not obviously distinguishable from
fluctuations, our data at high $1/B$ include $\mean{\lambda}$ values averaged
over such instabilities.}.


\section{Fluctuations}

\begin{figure}[tb]
\centerline{\includegraphics[width=6truein]{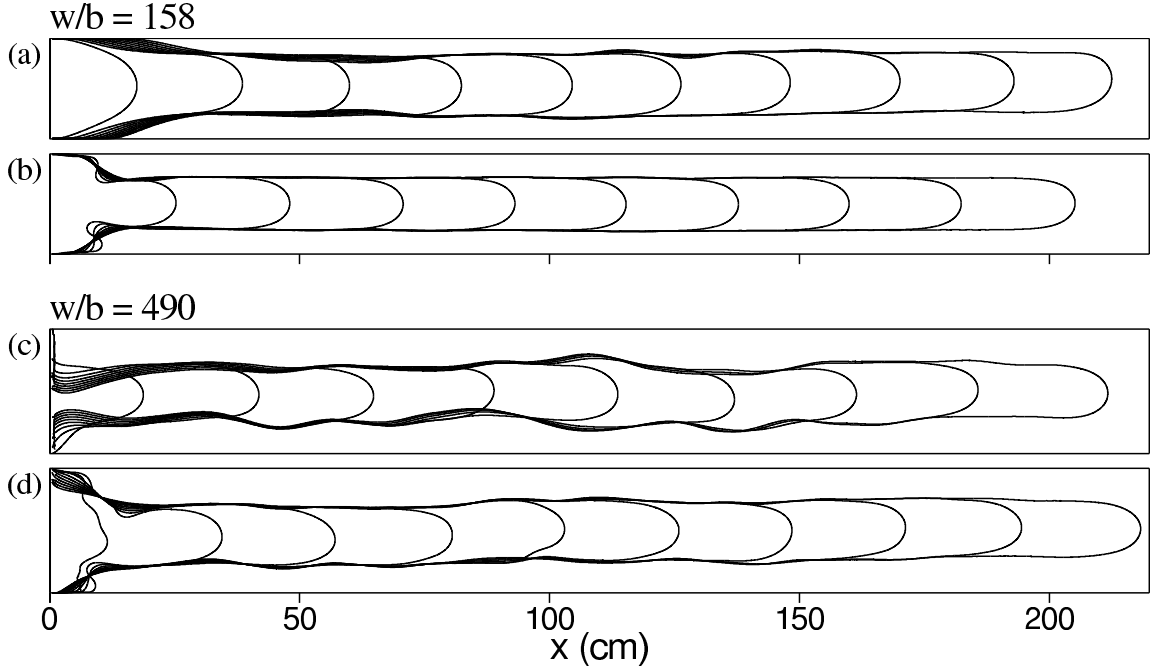}}
\caption{Finger images recorded at regular time intervals for different values
of aspect ratio $w/b$ and modified capillary number $1/B$.  Experimental values
for each run:  (a) $1/B = 116$, the rms fluctuation of the finger width
$\delta_\lambda = 1.30\times10^{-2}$, time between successive curves $\Delta t =
270$;  (b) $1/B = 1757$, $\delta_\lambda = 3.96\times10^{-3}$, $\Delta t = 18$;
(c) $1/B = 408$, $\delta_\lambda = 3.32\times10^{-2}$, $\Delta t = 810$;  (d)
$1/B = 2869$, $\delta_\lambda = 1.26\times10^{-2}$, $\Delta t = 108$.}
\label{fig:time_series}
\end{figure}

\begin{figure}[tb]
\centerline{\includegraphics[width=6truein]{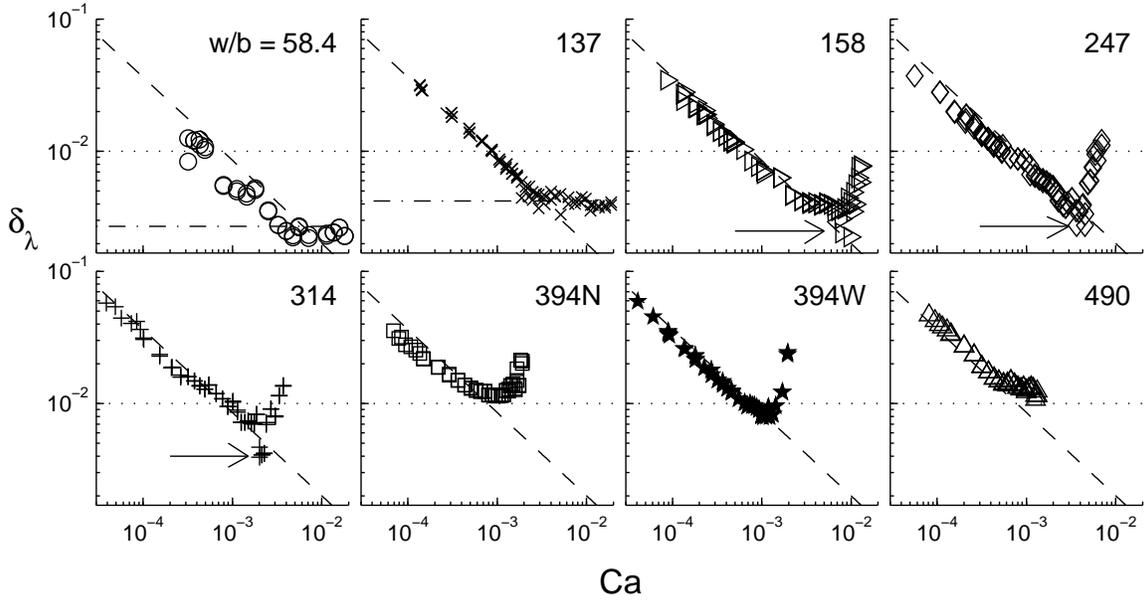}}
\caption{The rms fluctuation of the relative finger width $\delta_{\lambda}$ as
a function of capillary number Ca, where the dashed lines describe
the best fit to all of the data sets within the observed scaling
region, $\delta_{\lambda} = (\scint{1.1}{-4})\mathrm{Ca}^{-0.64}$.
(The data for 394N and 394W have the same aspect ratio but
different widths; the channel for 394N was 20.0~cm wide, 394W was 25.0~cm.)
The first two graphs are for data taken in the smaller channel; the
horizontal dash-dotted lines correspond to the limits of
measurement accuracy for that channel. Fluctuations with
$\delta_\lambda \lesssim 10^{-2}$ are not obvious visually, which may explain
why previous experiments at lower aspect ratios did not observe them. An
upturn in $\delta_\lambda$ occurs at high Ca, signalling the onset
of the secondary instabilities in the tips of the fingers.  The
arrows in the graphs for $w/b = 158$, 247, and 314 point to data from runs
where pinch-off events (see fig.~\ref{fig:pinch_off}) occurred, demonstrating
the delay of the onset of secondary instabilities in these cases.}
\label{fig:fluct_plot}
\end{figure}

Typical interface image sequences are shown for $w/b=158$ and 490 in
Fig.~\ref{fig:time_series}.  The finger width $\lambda$ (relative to the channel
width $w$) fluctuates visibly at low flow velocities for both aspect ratios
(Fig.~\ref{fig:time_series}~(a),(c)).  In the smaller aspect ratio system the
width appears to become steady as the finger velocity is increased
(Fig.~\ref{fig:time_series}(b)), appearing exactly like the classic ``half-width
finger'' of Saffman and Taylor. However, with sufficient resolution,
fluctuations can still be measured for all velocities up to the onset of tip
instabilities. In the higher aspect ratio system the width fluctuates visibly
for all flow rates (Fig.~\ref{fig:time_series}(d)). We find that the rms
fluctuation of the finger widths (relative to the channel width $w$) is
described by $\delta_\lambda = A(\mathrm{Ca})$${}^{\beta}$ with $A = (1.1 \pm
0.3) \times 10^{-4}$ and $\beta = -0.64 \pm 0.04$, for all the aspect ratios
studied (Fig.~\ref{fig:fluct_plot}).

The fluctuations in finger width are accompanied by a substantial deviation from
the expected relation between finger width and
velocity~\cite{Moore/Juel/etc:2002}.  For low aspect ratios, our finger
width measurements are in accord with previous results, but
for aspect ratios $w/b \gtrsim 250$, which were not examined in
previous work, we find that the mean finger width exhibits a surprising maximum
as the tip velocity is decreased.  However, the maximum value of the
fluctuating finger width observed during finger evolution,
$\lambda_{\mathrm{max}}$, still exhibits the
classical scaling with $1/B$ for all aspect ratios. This suggests that the
fluctuations represent an intermittent narrowing of the fingers from their
``ideal'' width~\cite{Moore/Juel/etc:2002}.


The finger width fluctuations and the peak in $\mean{\lambda}$
versus $1/B$ have proven robust under many variations of
experimental conditions, allowing us to rule out many possible
explanations for the phenomena.  We obtained identical results
with differences in the oil viscosity, the pinning properties at
the back of the channel, the pumping method, and the magnitude of
the channel's intrinsic gap variations~\cite{Moore/Juel/etc:2002}.
An instability of the film wetting layer is also unlikely, because
the film is very thin at low capillary numbers, where the
fluctuations are largest.  Though the fluctuation power law we
observed remained unchanged for {\it all} experimental variations,
we did discover that measurements on two channels with $w/b=394$
but different values of $w$ and $b$ yielded slightly different
results~\cite{Moore/Juel/etc:2002}. This difference suggests that
either the finger width for a given $w/b$ and Ca is not unique, or
perhaps a third parameter, as yet unknown, is necessary to fully
describe the problem.

\section{Pinch-off Events}

\begin{figure}[tb]
\centerline{\includegraphics[width=6truein]{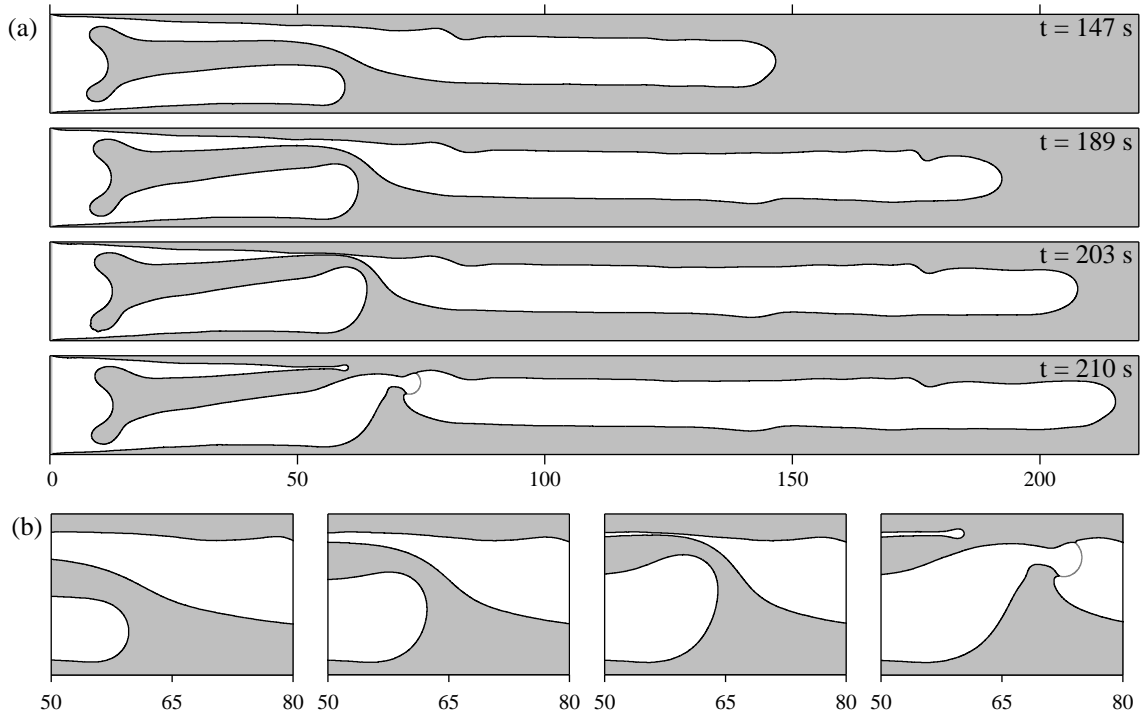}}

\caption{(a) A time sequence from a run where a pinch-off event
occurred; the times since the beginning of the pumping are shown.
The back of the growing finger narrows while the adjacent
suppressed finger grows.  The process accelerates as it continues;
the pinch-off and subsequent reconnection happens in less than two
seconds (which is faster than the time resolution we had for this data).  The
gray line in the last interface shows where the interface between the middle
finger and the disconnected bubble finally broke; this point could be clearly
seen on the image as a thick film left behind at that point.
(b) A close up view of the pinch-off at the same times. (The distance scales
shown are in cm.)
} \label{fig:pinch_off}
\end{figure}

In addition to finger width fluctuations, we have observed another unexpected
phenomenon: finger pinch-off events followed by reconnection to a different
finger, an example of which is shown in fig.~\ref{fig:pinch_off}.  For
sufficiently high flow rates, several fingers compete during the early stages
until one gets ahead, suppressing the growth of the other initial fingers.
However, if the initial competition stage was long, the suppressed fingers will
continue to interact with the active finger.  As the active finger continues to
grow, it narrows at the back, and the adjacent short finger grows towards it.
This process is very slow at first and gradually accelerates; the final
pinch-off and the connection of the previously suppressed finger to the
resulting bubble occurs very quickly.

The dynamics of the finger tip appear unaffected during these events, though the
onset of secondary instabilities is apparently slightly suppressed.  This can be
seen in fig.~\ref{fig:fluct_plot}; all of the data points indicated by arrows
had pinch-off events, and they continue to follow the fluctuation power-law when
other runs at similar flow rates without pinch-off events depart that curve,
beginning the upturn that marks the onset of the secondary instabilities.  At
higher flow rates, pinch-offs at the back of fingers and tip instabilities both
occur, apparently independently of one another.

Pinch-offs may or may not occur on a particular run at a given flow rate, since
this depends on the configuration of the initial finger competition, which can
be different each time.  In particular, the levelling of the cells is crucial: a
small tilt around the long axis will give fingers on the higher side of the
cell an early advantage; the shorter dormant fingers resulting from this do not
interact as quickly.

Pinch-off events are not mentioned in any of the previous literature; we believe
that we observe it because our cell has an unusually high ratio of length to
width, allowing sufficient time for evolution towards the pinch-off to occur
before the finger reaches the end of the channel.  Also, seemingly unimportant
asymmetries across the width of the cell, like the levelling mentioned above,
can disrupt the long period of finger competition and thus suppress the
pinch-off events.  We have observed the pinch-off phenomenon in different cells
at both high and low aspect ratios.  These events can occur at flow rates where
the plate bending due to the imposed pressure gradient is unmeasurably small.
The cause of this phenomenon is still unknown.


\section{Conclusions}

In conclusion, we have discovered fluctuations that intermittently
narrow evolving single fingers; the magnitude of these
fluctuations follows a power law with the capillary number for all
aspect ratios studied.  This is accompanied by a departure from
the classic scaling of finger width versus $1/B$ for large aspect
ratios ($w/b \gtrsim 250$).  In addition, we observe finger
pinch-off and reconnection events at the back of the channel;
these are apparently due to an interaction between adjacent
fingers that has not yet been explored.  Neither the fluctuation
nor the pinch-off phenomena are predicted by existing viscous
fingering theories, nor have they been observed in simulations.


\begin{theacknowledgments}
We thank J.\ B.\ Swift for frequent guidance and advice.  This work was funded
by the Office of Naval Research.
\end{theacknowledgments}

\bibliographystyle{aipproc}
\bibliography{refs}

\end{document}